\documentclass[aps,twocolumn,raggedbottom,showpacs,nobalancelastpage,amssymb,superscriptaddress]{revtex4}

\usepackage[english]{babel}
\usepackage{graphicx}
\usepackage{amssymb}
\usepackage{amsmath}
\usepackage{amscd}
\usepackage{eucal}
\usepackage{color}
\usepackage{bm}

\def\be{\begin{equation}}
\def\ee{\end{equation}}
\def\bea{\begin{eqnarray}}
\def\eea{\end{eqnarray}}
\def\bsub{\begin{subequations}}
\def\esub{\end{subequations}}

\usepackage{ulem}

\begin{document}

\title{Graphene-based heterojunction between two topological insulators}

\newcommand{\spsms}{CEA-INAC/UJF Grenoble 1, SPSMS UMR-E 9001, Grenoble F-38054, France}
\newcommand{\lyon}{CNRS - Laboratoire de physique, Ecole Normale sup\'erieure de Lyon, France}

\author{Oleksii Shevtsov}
\affiliation{\spsms}
\author{Pierre Carmier}
\affiliation{\spsms}
\author{Cyril Petitjean}
\affiliation{\spsms}
\affiliation{\lyon}
\author{Christoph Groth}
\affiliation{\spsms}
\author{David Carpentier}
\affiliation{\lyon}
\author{Xavier Waintal}
\affiliation{\spsms}
\date{\today}

\begin{abstract}
Quantum Hall (QH) and quantum spin Hall (QSH) phases have very different edge states and, when going from one phase to the other, the direction of one edge state must be reversed. We study this phenomenon in graphene in presence of a strong perpendicular magnetic field on top of a spin-orbit (SO) induced QSH phase. We show that, below the SO gap, the QSH phase is virtually unaffected by the presence of the magnetic field. Above the SO gap, the QH phase is restored. An electrostatic gate placed on top of the system allows to create a QSH-QH junction which is characterized by the existence of a spin-polarized chiral state, propagating along the topological interface. We find that such a setup naturally provides an extremely sensitive spin-polarized current switch.
\end{abstract}


\maketitle

\section{Introduction}

Electronic properties of graphene and topological insulators have received considerable attention these last few years \cite{DasSarma11,Hasan10}. Graphene is a two-dimensional crystal whose electronic band structure is that of a gapless semiconductor, with conduction and valence bands touching each other at two inequivalent points K and K' commonly referred to as valleys. The energy of charge carriers vanishes at these points and disperses linearly with momentum in their vicinity, forming a so-called Dirac cone. Low-energy excitations are massless Dirac fermions. These exotic quasi-particles carry a topological Berry phase which has been shown to give rise to many unusual transport phenomena such as the suppression of backscattering (also known as Klein tunnelling) \cite{Ando98,Young09}, weak anti-localization \cite{Suzuura02,Tikhonenko09}, and a ``relativistic" quantum Hall effect \cite{Novoselov05,Zhang05,Gorbig11}. Some of these properties, however, are not robust to the presence of disorder -- as soon as the latter is sufficiently short-ranged to induce valley-mixing -- due to the existence in graphene of an even number of Dirac cones.

\begin{figure}[]
\begin{center}
\includegraphics[angle=0,width=0.9\linewidth]{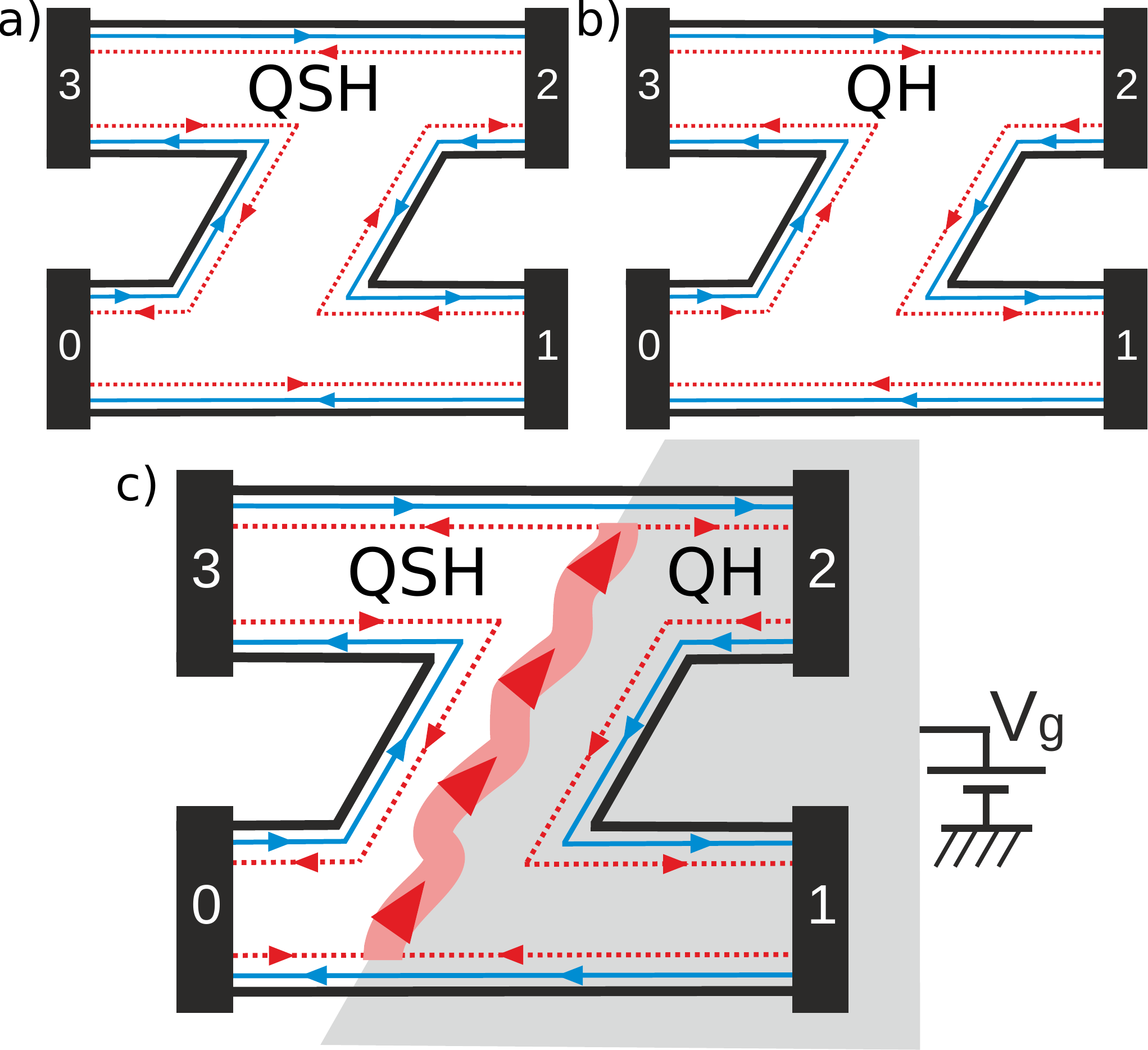}
\caption{Cartoon for the edge states in a 4-terminal Z-shape geometry when the system is in the \textbf{a)} QSH phase and the \textbf{b)} QH phase. In the former case, opposite spins (thick blue and red dashed lines) propagate in opposite directions on a given edge, while in the latter case they propagate in the same direction. \textbf{c)} Topological heterojunction, with QSH edge states on the left and QH edge states on the right. This junction can be experimentally achieved by applying a top gate (grey shaded region) on the right half of the sample. While one of the spin species (thick blue) can propagate through this junction, the other one (red dashed) cannot and therefore gives rise to a chiral state, localized at the interface between QH and QSH phases, which connects both edges.}
\label{Fig1}
\end{center}
\end{figure}

On the other hand, topological insulators are a newly discovered type of system which are insulating in the bulk and characterized by the existence of robust gapless excitations at their surface \cite{Hasan10,Kane05bis,Kane05,Bernevig06,Moore07,Roy09}.
In two dimensions, the topologically insulating phase possesses gapless states propagating along its edges. There are two famous examples of this phase: the quantum Hall (QH) insulator, which can be obtained by applying a strong magnetic field perpendicular to the plane and is characterized by chiral spin-degenerate edge states (Fig.~\ref{Fig1}b), and the time-reversal-symmetric quantum spin Hall (QSH) insulator, which is induced by a strong spin-orbit (SO) interaction \cite{Kane05bis,Bernevig06,Konig07} and is characterized by states with opposite spins propagating in opposite directions (Fig.~\ref{Fig1}a).

Whereas examples of three-dimensional topological insulators have been found to abound in Nature, two-dimensional systems exhibiting the QSH phase are so far limited to HgTe/CdTe heterostructures which only a few experimental groups in the world can synthetize. It was recently suggested \cite{Weeks11} that graphene might come to the rescue, much like it did with two-dimensional electron gases which were until a few years ago limited to epitaxially grown semiconducting heterostructures. Graphene's versatile fabrication methods and insensitivity to ambient temperature and chemical conditions have indeed raised important expectations for all sorts of applications in electronics \cite{Geim09}. Following the seminal observation by Kane and Mele that graphene with SO coupling should be a QSH insulator \cite{Kane05bis}, Weeks et al. numerically predicted the possibility of opening a substantial QSH gap in graphene by depositing on it heavy adatoms (such as In or Tl) which can locally induce strong SO coupling in the system. In a previous paper \cite{Shevtsov11}, we studied in detail the transport in graphene randomly covered by diluted adatoms and showed that the QSH phase was insensitive to the inhomogeneity in the coverage $n_{\text{ad}}$ and that, for all purposes, the system could be mapped to a homogeneous phase with a renormalized SO coupling strength $\lambda_{\text{so}}^{\text{eff}}=\lambda_{\text{so}} n_{\text{ad}}$. We explore in this article the fate of the QSH phase when a strong perpendicular magnetic field is applied to graphene. Surprisingly, we show that the QSH phase is preserved for energies below the QSH gap, even for extremely strong magnetic fields or in the presence of disorder. As we will see in the following, this can be traced back to the peculiar "relativistic" aspect of the QH effect in graphene (as well as to the negligible contribution of Zeeman splitting in graphene). The often quoted simplified picture of SO coupling acting as an effective magnetic field with opposite signs for opposite spins clearly breaks down in this situation. Next, we turn to the investigation of the transport properties of a junction between a QSH and a QH phase (Fig.~\ref{Fig1}c). We show that this setup features a robust state, localized at the interface between the two topological insulators, analogous to the ambipolar ``snake" states which arise in graphene quantum Hall $n$-$p$ junctions \cite{Williams11}, and take advantage of it by demonstrating how it can serve to realize a topologically protected spin-polarized charge-current switch.

The paper is organized as follows. In section \ref{sec:Mot}, we recall the basic features of the QH and QSH phases and introduce the physical question at the root of this work. In section \ref{sec:Band}, we discuss the band structure of our system and show in section \ref{sec:Top} how the use of topological Chern numbers can provide us with a clean and simple interpretation for it. Numerical transport calculations are presented in section \ref{sec:Trans}, in which we show that all regimes (below, above, and at the topological transition) possess characteristic signatures. The effects of geometry and disorder are discussed. Transport in the presence of a heterojunction between our two topologically insulating phases is analyzed in section \ref{sec:Het}. We discuss similarities with other setups such as quantum Hall $n$-$p$ junctions, and comment on the possibility of realizing a spin-polarized current switch with huge on/off ratios. Our results are summarized in section \ref{sec:Conc}.

\section{How to reverse the direction of propagation of an edge state ?}
\label{sec:Mot}

The issue of how time-reversal-symmetry breaking can affect the QSH phase has been addressed previously in the literature in different settings \cite{Tkachov10,Yang11,DeMartino11,Goldman11}. To the best of our knowledge, however, transport signatures of the competition between QH and QSH phases in graphene have not been considered as of yet, with the exception of the work by Abanin et al. \cite{Abanin06} where the QSH phase arose from a different mechanism (Zeeman splitting), which is extremely weak\,\footnote{It is so weak that it appears to be superseded by many-body effects in strong magnetic fields \cite{Zhang06,Jiang07}. If Zeeman splitting $\epsilon_Z$ were artificially enhanced, it would compete with SO coupling and lead to a phase transition when $\epsilon_Z=\Delta_{\text{so}}$, characterized by a reversal of the direction of spin current on a given edge \cite{DeMartino11}.} and can therefore easily be destroyed by local fluctuations of the magnetic field. In contrast, we study the model introduced by Kane and Mele \cite{Kane05bis}, to which we add the presence of a strong perpendicular magnetic field:
\be
\label{eq:KMB}
H = v_F(\hat{\Pi}_x\sigma_x\tau_z+\hat{\Pi}_y\sigma_y)+\Delta_{\text{so}}\sigma_z\tau_zs_z \; .
\ee
$\hat{{\bf \Pi}}=\hat{{\bf p}}+e{\bf A}$ is the generalized momentum which accounts for the presence of the magnetic vector potential ${\bf A}$ associated with a perpendicular magnetic field ${\bf B}=B{\bf z}$ ($\nabla\times{\bf A}={\bf B}$), $\Delta_{\text{so}}$ is the SO-induced QSH gap, and $v_F=3ta_{C-C}/(2\hbar)$ is the Fermi velocity  (expressed as a function of the microscopic lattice parameters $t$ (nearest-neighbor hopping amplitude) and $a_{C-C}$ (nearest-neighbor distance) which we choose in the following as our working units of energy and length, respectively). $\{\sigma,\tau,s\}$ are Pauli matrices in, respectively, sublattice, valley and spin spaces.

\begin{figure}[]
\begin{center}
\includegraphics[angle=0,width=1.0\linewidth]{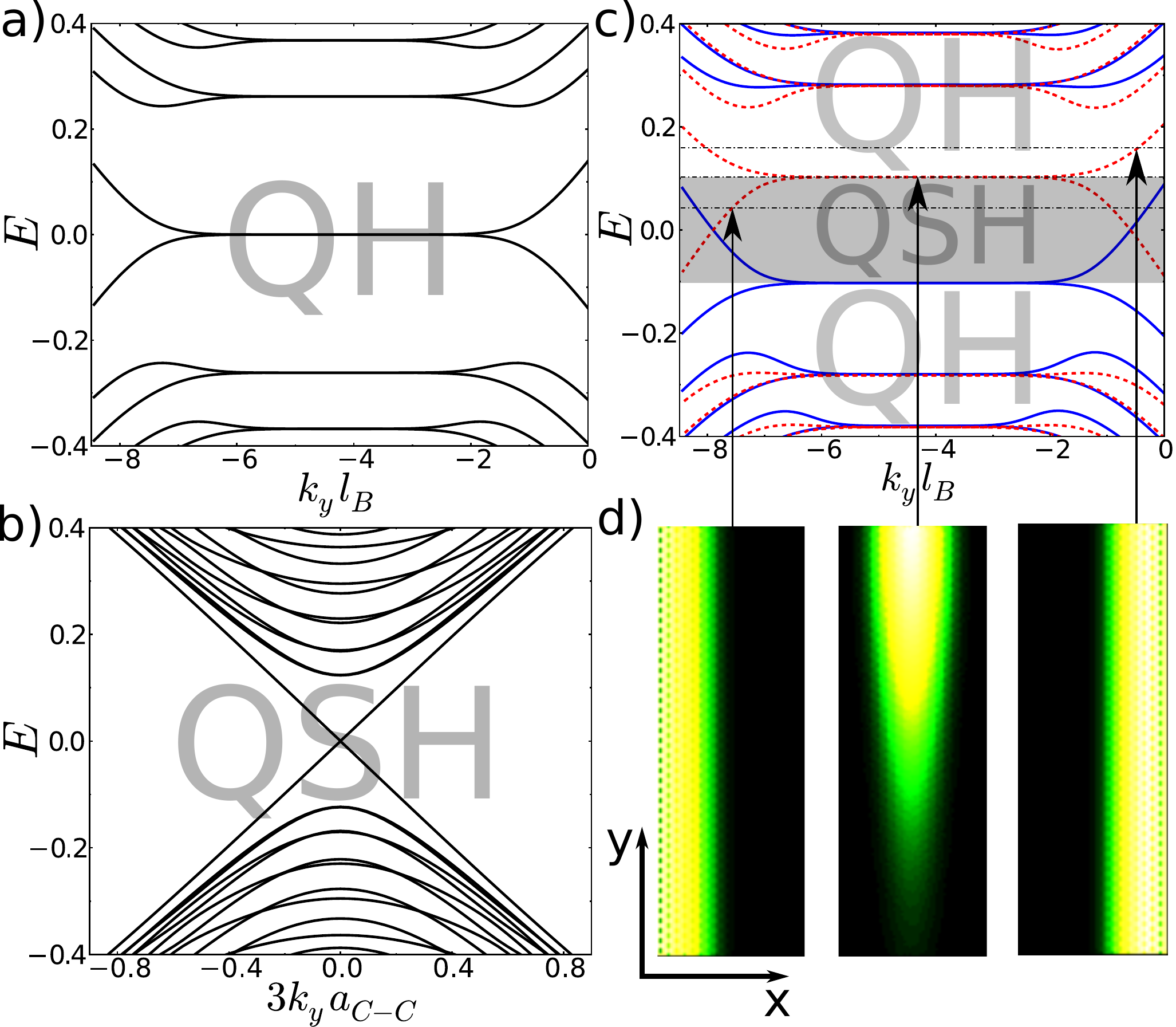}
\caption{Band structure of a (semi-metallic armchair) graphene ribbon in the \textbf{a)} QH and \textbf{b)} QSH phase. When both magnetic field and SO coupling are present \textbf{c)}, the resulting band structure leads to a QSH phase for $|E_F|<\Delta_{\text{so}}$ (shaded region) and a QH phase for $|E_F|>\Delta_{\text{so}}$. Compared to the pure QH and QSH cases, the spin degeneracy is lifted (blue thick and red dashed lines), which is particularly prominent in the lowest band which consists of spin-polarized branches at $E=\pm\Delta_{\text{so}}$. As the Fermi energy crosses the SO gap, the localization of the ``unhappy" spin (red dashed) shifts from one edge to the other, while it is fully localized in the bulk when $E_F=\Delta_{\text{so}}$. This is illustrated in the corresponding current-density plots \textbf{d)}. On the other hand, the ``happy" spin (thick blue) gets increasingly localized on the same edge as the Fermi energy crosses the transition region (not shown). The parameters used are $\lambda_{\text{so}}=0.02$, $l_B\simeq8$, and a ribbon width $W=40\sqrt{3}$.}
\label{Fig2}
\end{center}
\end{figure}

The band structures of a graphene ribbon in the QSH and QH phases are quite different. In the QH phase,  the perpendicular magnetic field gives rise to Landau levels $\epsilon_n=\pm(\hbar v_F/l_B)\sqrt{2|n|}$, with $l_B=\sqrt{\hbar/(eB)}$, which become dispersive close to the edges of the graphene ribbon (Fig.~\ref{Fig2}a). When the Fermi level is placed between two of these Landau levels, transport is characterized by spin-degenerate edge states as in Fig.~\ref{Fig1}b, which propagate in a direction imposed by the sign of the magnetic field. In the QSH phase, the band structure consists of hyperbolic bands above the QSH gap and a pair of linearly dispersing ones below it (Fig.~\ref{Fig2}b). These linear bands correspond to spin-polarized states, localized at the edges of the graphene ribbon on a characteristic length scale $\xi_{\text{so}}=\hbar v_F/\Delta_{\text{so}}$. When the Fermi level is below the QSH gap, transport in the system can be described by counter-propagating edge states as shown in Fig.~\ref{Fig1}a. By comparing Figs.~\ref{Fig1}a and \ref{Fig1}b, one observes that one spin species -- hereafter referred to as the ``unhappy" spin (dashed red in Fig.~\ref{Fig1}) -- has to reverse its direction of propagation when going from one phase to the other. This article is devoted to the study of how this reversal happens in two different setups: a homogeneous sample, where the quantum phase transition between QH and QSH phases is driven by electrostatic doping, and a heterojunction between the two phases.

\section{Band structure}
\label{sec:Band}
The first insight is given by the spectrum of Eq.~\ref{eq:KMB} in the presence of graphene edges. The energy spectrum for a bulk system described by Eq.~\ref{eq:KMB} reads $\epsilon_n=\pm\sqrt{\Delta_B^2|n|+\Delta_{\text{so}}^2}$, with $\Delta_B=(\hbar v_F/l_B)\sqrt{2}$. The lowest level $n=0$ stands out from the others, as each branch ($\pm$) can be shown to host only one of the two spin species \cite{DeMartino11} (Fig.~\ref{Fig2}c). Whereas $n>0$ levels will all disperse in the same direction when confinement is taken into account, the fate of the lowest level is more subtle.
To be more quantitative, we make use of a tight-binding model on the graphene hexagonal lattice, which in the presence of both Kane-Mele SO coupling and perpendicular magnetic field can be written as
\be
\label{eq:tb}
H = -t\sum_{\langle i,j \rangle,\alpha} e^{i\phi_{ij}} c_{i,\alpha}^\dagger c_{j,\alpha}
+ i\lambda_{\text{so}}\sum_{ \langle\langle i,j\rangle\rangle,\alpha,\beta}\nu_{ij}e^{i\phi_{ij}} c_{i,\alpha}^\dagger s_{\alpha\beta}^z c_{j,\beta} \; .
\ee
Indices ($i,j$) label lattice sites, ($\alpha,\beta$) label spin indices, while symbols $\langle\;\rangle$ and $\langle\langle\;\rangle\rangle$ respectively refer to nearest-neighbor coupling (with hopping amplitude $t$) and next-nearest-neighbor coupling (with SO-induced hopping amplitude $\lambda_{\text{so}} = \Delta_{\text{so}}/(3\sqrt{3})$ \cite{Kane05bis}). The Peierls phase $\phi_{ij}=(e/\hbar)\int_{{\bf r}_j}^{{\bf r}_i}{\bf A}\cdot d{\bf r}$ takes into account the contribution from the magnetic flux threading the lattice, and $\nu_{ij}=\pm1$ depending on whether sites are coupled clockwise or counter-clockwise. Note that in order for the system to remain gauge invariant, Peierls substitution has to be done on all hopping matrix elements: nearest-neighbor {\it and} (SO) second nearest-neighbor. To compute transport properties numerically, we make use of the software KNIT, which is based on an advanced recursive Green's function technique \cite{Kazymyrenko08}, and works in the linear response regime. The numerical calculations are done with semi-metallic armchair boundary conditions, but our results are qualitatively unaffected by this choice. An important technical point is that the magnetic field should be present in the entire sample, including in the leads, in order to avoid spurious reflection at the lead-sample interface. The multi-terminal Peierls subsitution prescription allowing to do that is described in the Appendix.

The full tight-binding band structure of a semi-metallic armchair graphene ribbon described by the Hamiltonian in Eq.~(\ref{eq:tb}) is shown on Fig.~\ref{Fig2}c. It can be summarized in very simple terms: for Fermi energies inside the SO gap $|E_F|<\Delta_{\text{so}}$ (shaded region), the system is in the QSH phase, with opposite spin channels on a given edge propagating in opposite directions, while for energies $|E_F|>\Delta_{\text{so}}$, the system is in the QH phase, with opposite spin channels on a given edge propagating in the same direction. Hence for a given value of $\Delta_{\text{so}}$, the transition between the two phases is governed solely by the Fermi energy and does not depend at all on the value of the magnetic field (once again neglecting Zeeman splitting, which is very small in graphene).
This quite remarkable result is a direct consequence of the existence in graphene of a $B$-independent zero-energy Landau level: as soon as $\Delta_{\text{so}}\neq0$, the spin degeneracy of the zero-energy Landau level is lifted, as opposed to all other Landau levels which remain spin degenerate \cite{DeMartino11}. This lifting leads to a QSH phase in the corresponding SO gap, as can be understood with the help of topological invariants.

\section{Topological order}
\label{sec:Top}

In this section, we discuss the topological order of the phases obtained by varying the chemical potential in the energy spectrum of Fig.~\ref{Fig2}c. In particular, we relate the unique transition between the QSH topological order and the QH topological order in presence of SO coupling to the specificities of the QH physics of Dirac fermions in graphene. Let us start by recalling the standard topological number characterization of Landau levels when $\Delta_{\text{so}} = 0$. Each Landau level $n$ and its associated eigenfunctions over the first Brillouin zone are characterized by a topological invariant, the so-called Chern number \cite{Thouless82}. This topological number takes a value ${\cal C}^{(n)}_{\tau,s}=+1$ for each Landau level, independently of the Landau $n$, valley $\tau$
or spin $s$ indices. For each value of the Fermi energy, we can characterize the corresponding phase by a topological number ${\cal C}=\sum_{\tau,s}\sum_{\epsilon_n<E_F}{\cal C}^{(n)}_{\tau,s}$ obtained by summing the Chern numbers of all filled energy bands \cite{Thouless82}. For graphene and any Dirac system, however, this procedure would yield an ill-defined topological number ${\cal C}$ due to the presence of an infinite number of filled Landau levels below $E_{F}$. As shown recently through the use of non-commutative Berry's connection \cite{Watanabe11}, the correct topological number for a single Dirac cone takes a value ${\cal C}_{\tau,s} = -1/2$ for $\epsilon_{-1}<E_{F}<\epsilon_{0}$, and ${\cal C}_{\tau,s} = +1/2$ for $\epsilon_{0}<E_{F}<\epsilon_1$, where $\epsilon_n$ are the usual Landau levels and ${\cal C}_{\tau,s}=\sum_{\epsilon_n<E_F}{\cal C}^{(n)}_{\tau,s}$. In the discussion below, we will make use of the more convenient topological Chern number per spin species, ${\cal C}_{s} = \sum_{\tau} {\cal C}_{\tau,s} $ which takes values two-fold larger.

Let us now turn to the energy spectrum of Fig.~\ref{Fig2}c where $\Delta_{\text{so}} \neq 0$. The presence of the SO coupling does not modify any of the Chern numbers per Landau level, but it lifts the spin degeneracy of the $n=0$ Landau level into the two levels at $E=\pm \Delta_{\text{so}}$. As the $z$ component of spin is conserved, the topological Chern numbers per spin species introduced above turn out to be useful quantities to characterize the topological order in this new spectrum. They read
\begin{subequations}
\label{eq:spinChern}
\begin{align}
 &{\cal C}_{\uparrow} = -1,  {\cal C}_{\downarrow} = -1 \;
 \textrm{ for }\epsilon_{-1}<E_{F}<-\Delta_{\text{so}} \;, \\
&{\cal C}_{\uparrow} = +1,  {\cal C}_{\downarrow} = -1 \;
 \textrm{ for }-\Delta_{\text{so}}<E_{F}<\Delta_{\text{so}} \;, \\
&{\cal C}_{\uparrow} = +1,  {\cal C}_{\downarrow} = +1 \;
 \textrm{ for }\Delta_{\text{so}}<E_{F}<\epsilon_1 \;,\\
 &{\cal C}_{\uparrow} = +3,  {\cal C}_{\downarrow} = +3 \;
 \textrm{ for }\epsilon_1<E_{F}<\epsilon_2 \;,
 \end{align}
 \end{subequations}
this time with $\epsilon_n$ the modified Landau levels introduced at the beginning of section \ref{sec:Band}. The difference between ${\cal C}_{\uparrow}$ and ${\cal C}_{\downarrow}$ for $|E_{F}|<\Delta_{\text{so}}$ signals the appearance of a  $\mathbb{Z}_{2}$ topological order characteristic of the QSH phase. Indeed, when the $z$ component of spin is conserved, the $\mathbb{Z}_{2}$ topological index characterizing the QSH phase is defined as $\nu=({\cal C}_{\uparrow} -{\cal C}_{\downarrow})/2$ (mod $2$) \cite{Sheng06,Hasan10}. As all Landau levels $n\neq0$ are still spin-degenerate, we have ${\cal C}_{\uparrow} = {\cal C}_{\downarrow} $ and thus $\nu=0$ for all Fermi energies $|E_{F}|>\Delta_{\text{so}}$. For these values of $E_{F}$, the system lies in a QH phase characterized by the usual topological Chern number ${\cal C}={\cal C}_{\uparrow} + {\cal C}_{\downarrow}$. However for $|E_{F} | < \Delta_{\text{so}}$, Eq.~(\ref{eq:spinChern}) leads to a non-trivial $\mathbb{Z}_{2}$ index $\nu = 1$, while the total Chern number simultaneously vanishes ${\cal C}={\cal C}_{\uparrow} + {\cal C}_{\downarrow}=0$. The system then lies in a different topologically insulating phase: the QSH insulator. This shows that as the Fermi energy crosses the values $\pm \Delta_{\text{so}}$, the system undergoes a quantum phase transition between two topological insulators: a QH phase and a QSH phase:
\begin{subequations}
\label{eq:spinChern2}
\begin{align}
 &\nu = 0,  {\cal C} = -2 \;
 \textrm{ for }\epsilon_{-1}<E_{F}<-\Delta_{\text{so}} \;: \textrm{ QH } \;, \\
&\nu = 1,  {\cal C} = 0 \;
 \textrm{ for }-\Delta_{\text{so}}<E_{F}<\Delta_{\text{so}} \;\;: \textrm{ QSH } \;, \\
&\nu = 0,  {\cal C} = 2 \;
 \textrm{ for }\Delta_{\text{so}}<E_{F}<\epsilon_1 \;\;: \textrm{ QH } \;, \\
 &\nu = 0,  {\cal C} = 6 \;
 \textrm{ for }\epsilon_1<E_{F}<\epsilon_2 \;\;: \textrm{ QH } \;.
 \end{align}
 \end{subequations}
This transition appears crucially tied to the Dirac physics of graphene and the presence of the $n=0$ Landau level: in the present case, we do not need a SO coupling to overcome an energy gap in order to drive this transition, as would be the case for non-relativistic fermions with Landau gap $\hbar\omega_c$ or if graphene had a trivial mass gap $m\sigma_z$. The spin degeneracy lifting of the $n=0$ level is all that is required here. Let us note finally that while this argument formally uses the conservation of the $S_{z}$ spin component, the robustness of topological numbers proves it to remain valid if non-$S_{z}$-conserving terms are included in the Hamiltonian, provided two branches of the $n=0$ Landau level with opposite spins remain non-degenerate.

\section{Transport signatures}
\label{sec:Trans}

\subsection{Ballistic regime}

Let us now study how this topological phase transition appears in transport. We will study the multi-terminal (dimensionless) differential conductance $T_{ab}$
which expresses how much current $dI_a$ is collected in lead $a$ when the voltage in lead $b$ is raised by $dV_b$,
\be
\frac{dI_a}{dV_b}=\frac{e^2}{h} T_{ab} \; .
\ee
Additionally, in order to observe the edge states directly, we also study (in color plots) the differential local current density $di(\vec r)/dV_a$ which allows to clearly observe the
edge states inside the sample.
In Fig.~\ref{Fig2}d, we show local current-density plots which illustrate how the behavior of the ``unhappy" spin changes as a function of the Fermi energy: it goes from propagating along one edge when $E_F<\Delta_{\text{so}}$ (left panel) to propagating along the other as $E_F>\Delta_{\text{so}}$ (right panel), while it gets localized in the bulk at the critical point $E_F=\Delta_{\text{so}}$ (middle panel). We find that the energy window where this localization is observed is extremely narrow and decays exponentially with the width of the sample. Note that this scenario is completely different from what one would expect starting from the naive toy model of SO coupling acting as a spin-dependent magnetic field $B_{\text{so}}{\bf z}s_z$. In this case, for a critical value of the real magnetic field $B=B_{\text{so}}$, the ``unhappy" spin would feel no magnetic field at all and be fully delocalized. On the contrary, we observe that the QSH phase is virtually independent from the real magnetic field and that the ``unhappy" spin actually gets localized when $E_F=\Delta_{\text{so}}$, illustrating the limitations of the toy model in this situation.

In the vicinity of the transition, the ``unhappy" spin keeps propagating along a given edge but its classical cyclotron orbit center $x_c=-k_yl_B^2$ (with $k_y$ the longitudinal wave vector component) is shifted inwards as the Fermi energy increases (see Fig.~\ref{Fig2}c). On the other hand, the ``happy" spin gets increasingly localized on the same edge when the Fermi energy increases, which is qualitatively equivalent to the usual QH case. One can indeed show that the notion of a classical cyclotron orbit center remains well defined here, despite the presence of SO coupling, and that the corresponding eigenstates are very similar to those found in the pure QH regime \cite{DeMartino11}.

Before presenting the rest of our numerical data on transport in the vicinity of the QSH-QH transition, a few words should be said about the shape of the spectrum corresponding to the non-zero Landau levels. While the branches corresponding to the ``happy" spin seem basically unaffected by the SO coupling, the two branches of the ``unhappy" spin display very different behaviors (Fig.~\ref{Fig2}c). In particular, one of them is significantly bent by the SO coupling, such that counter-propagating states along the same edge appear in a finite window of energy. This leads to the possibility of backscattering and therefore destroys the robustness of the QH phase in this energy window, the size of which can nevertheless be significantly reduced by increasing the width of the graphene ribbon. Note that at negative energies, $E_F<-\Delta_{\text{so}}$, ``happy" and ``unhappy" spin species exchange their roles (see Fig.~\ref{Fig2}c).

\begin{figure}[]
\begin{center}
\includegraphics[angle=0,width=0.9\linewidth]{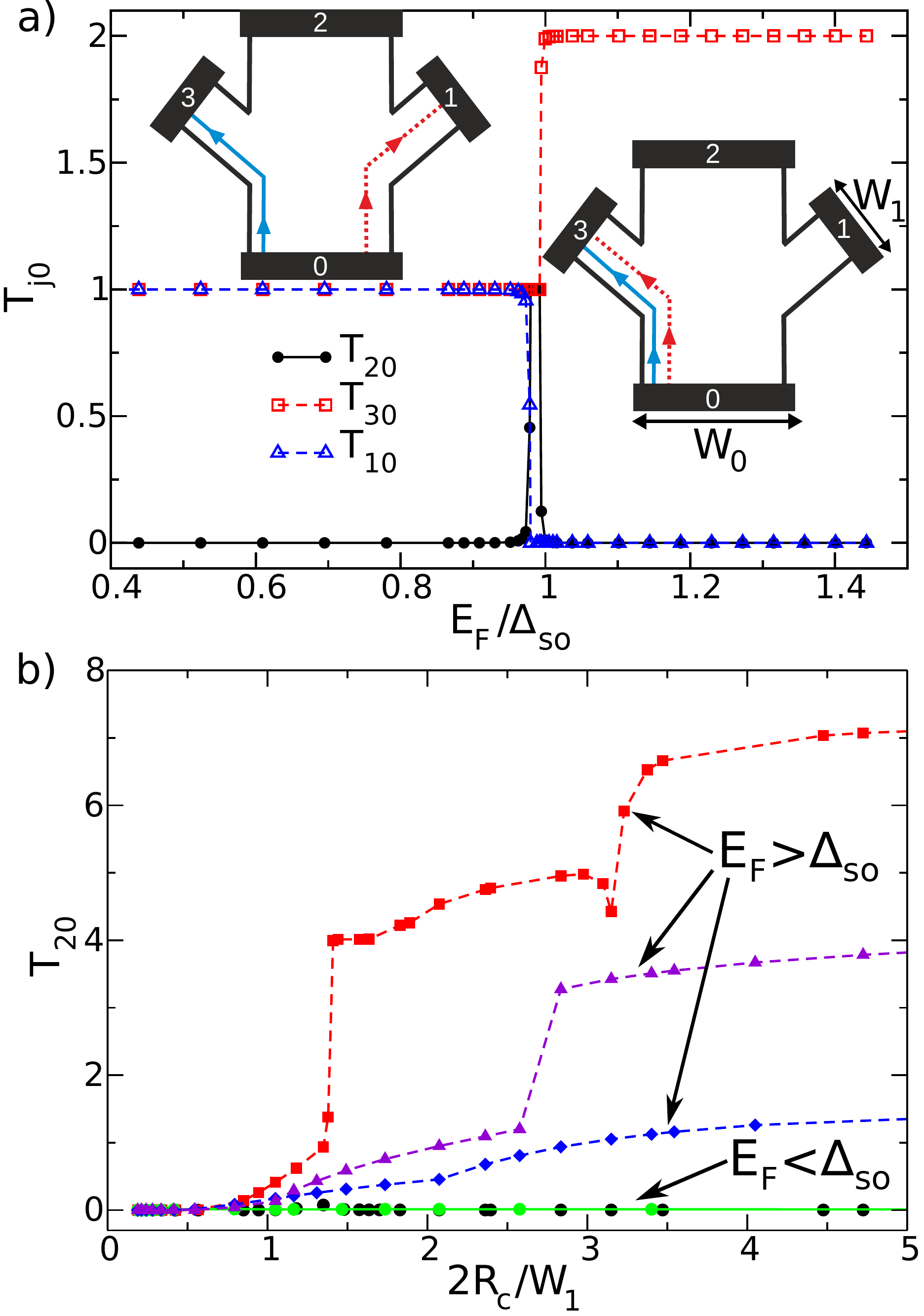}
\caption{QSH-QH transition in a 4 terminal $\Psi$-shape geometry. \textbf{a)} Dimensionless differential conductance from lead 0, where the current is injected, to outgoing leads 1, 2 and 3 in the geometry depicted in the inset, as a function of the Fermi energy ($\lambda_{\text{so}}=0.02$, $l_B\simeq8$, and lead widths $W_0=64\sqrt{3}$, $W_1=22\sqrt{3}$). Below the SO gap, current is carried by QSH edge states, which propagate on different edges (left inset), while above the SO gap it is carried by QH edge states, which propagate on the same edge (right inset). \textbf{b)} Dimensionless differential conductance from lead 0 to lead 2 as a function of the classical cyclotron radius $R_c$ (same values of $\lambda_{\text{so}}$, $W_0$ and $W_1$ as in \textbf{a}). Circular symbols (thick line) correspond to data with $E_F<\Delta_{\text{so}}$, while other symbols (dashed lines) correspond to data with $E_F>\Delta_{\text{so}}$, respectively
$E_F/\Delta_{\text{so}}=1.06$ (blue diamonds),
$E_F/\Delta_{\text{so}}=1.25$ (violet triangles), and
$E_F/\Delta_{\text{so}}=1.44$ (red squares). While this ``direct" transmission is vanishingly small and independent of the magnetic field below the SO gap, it increases with both $R_c$ and $E_F$ above the SO gap. The latter situation arises because the QH phase is destroyed as $2R_c\gtrsim W_1$, and because the number of transmitting channels increases with $l_B$ and $E_F$.}
\label{Fig3}
\end{center}
\end{figure}

We proceed to investigate further how the transition between QSH and QH phases shows up in transport, and consider the 4-terminal $\Psi$-shape geometry depicted in the inset of Fig.~\ref{Fig3}a. In the core of Fig.~\ref{Fig3}a, we plot as a function of the Fermi energy the current collected in leads 1, 2 and 3 when injected from lead 0. Nothing unexpected happens away from the transition: the transmission coefficients feature characteristic signatures of current-carrying edge states (left and right panels of Fig.~\ref{Fig2}d). Around the critical value $E_F=\Delta_{\text{so}}$, however, we observe in Fig.~\ref{Fig3}a that the ``unhappy" spin is fully transmitted in lead 2. This can be understood as follows. As the transition point is approached, the classical cyclotron orbit center is shifted (inwards) away from the edge (as discussed above) by a distance which can reach $2R_c$, where $R_c=l_B^2/\xi_{\text{so}}$ is the classical cyclotron radius of the $n=0$ level. When $2R_c>W_1/2$, where $W_1$ is the width of lead 1, then, somewhere in the vicinity of the transition, the incoming state cannot penetrate lead 1, hence leading to perfect direct transmission to lead 2 as observed in Fig.~\ref{Fig3}a. This feature could in principle be checked experimentally by simply varying the strength of the magnetic field close to the transition: for wider systems or at larger magnetic field, the width of the peak of direct transmission from 0 to 2 shrinks and eventually disappears.

A clear way to distinguish between QH and QSH phases in a transport measurement is to plot the ``direct" dimensionless differential conductance from lead 0 to lead 2 as a function of the classical cyclotron radius $R_c\propto B^{-1}$. This is illustrated in Fig.~\ref{Fig3}b: below the QSH gap, the ``direct" transmission is zero for any value of the magnetic field; on the other hand, above the QSH gap, transmission can become non-zero as the magnetic field weakens, due to the breakdown of the QH effect when $R_c$ becomes larger than the width of the system. It thus interpolates between zero (in the QH regime) and a finite value which depends on the number of occupied bands at the Fermi level in the ribbon when $B=0$. Of course, in the latter limit, the topological protection is lost and the value of the transmission will strongly depend on the disorder configuration.

\subsection{Effect of disorder}

In this section, we study the robustness of the results described in the ballistic regime with respect to the presence of various types of disorder. In the absence of a magnetic field, the QSH phase is known to be very robust with respect to the presence of scalar disorder (i.e. disorder that breaks neither time reversal symmetry nor spin rotational symmetry), as introducing backscattering between edge states involves tunneling through the gapped bulk region. On the other hand, disorder that breaks both time reversal symmetry and spin rotational symmetry
(such as magnetic impurities) leads to scattering between the two counter-propagating edge states of the same edge, which leads to intra-edge backscattering. If this intra-edge backscattering becomes strong (or the edges are very long), edge states may eventually get localized, such that no edge transport occurs anymore. In this paper, as we explicitly break time reversal symmetry with a magnetic field, disorder that breaks only spin rotational symmetry (such as Rashba-like SO terms arising from adatoms) effectively behaves as magnetic impurities and could potentially lead to the same breakdown of the QSH phase.

In order to study these effects quantitatively, we add the general perturbation 
\be
\label{eq:dis}
H_{\text{dis}} = \sum_{i,\alpha,\beta} c_{i,\alpha}^\dagger \left(\sum_{\mu=0,x,y,z} V_{i,\mu} s_{\alpha\beta}^\mu\right) c_{i,\beta}
\ee
to our tight-binding Hamiltonian (\ref{eq:tb}). The onsite potentials $V_{i,\mu}$ are independent random variables uniformly distributed inside a given interval on each site of the system (Anderson disorder). We study three different sorts of disorder with different symmetries: $V_{i,0}\in[-V_s/2,V_s/2]$ takes into account scalar (spin-independent) disorder, $V_{i,z}\in[-V_z/2,V_z/2]$ represents Zeeman-like ($S_z$ conserving) impurities, and $\{V_{i,x}, V_{i,y}\}\in[-V_m/2,V_m/2]$ captures the influence of $S_z$ non-conserving impurities.

\begin{figure}[]
\begin{center}
\includegraphics[angle=0,width=1.0\linewidth]{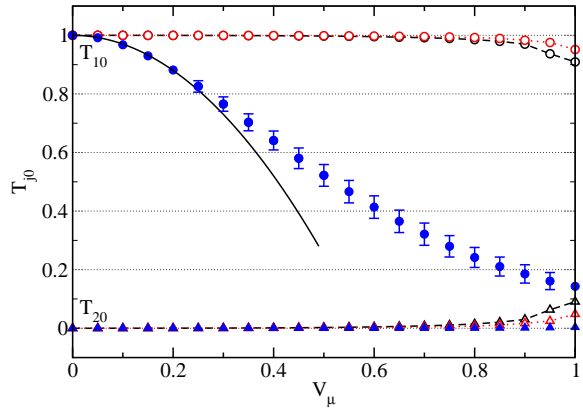}
\caption{Effect of very strong disorder on the QSH phase in the 4-terminal $\Psi$-shape geometry. We plot the dimensionless differential conductance from lead 0 to leads 1 and 2 as a function of disorder strength for $\lambda_{\text{so}}=0.02$, $l_B\simeq8$, $E_F=0.05$, and otherwise same system as in Fig.~\ref{Fig3}. Circular symbols stand for $T_{10}$ and triangular symbols for $T_{20}$. The behavior for $T_{30}$ (not shown) is the same as for $T_{10}$. Dashed (black) lines correspond to purely scalar disorder ($V_s$), dotted (red) lines correspond to purely Zeeman-like disorder ($V_z$), and filled (blue) symbols correspond to purely $S_z$ non-conserving disorder ($V_m$). The latter is clearly the dominant effect on $T_{10}$, though unrealistically large values are required to quantitatively affect the edge state transmission (see main text for a mean free path estimate). Each data point has been averaged over 48 disorder configurations. Unless shown, error bars are smaller than symbol sizes. Full (black) line: fit $y=1-3.0 x^2$.}
\label{Fig4}
\end{center}
\end{figure}

The results are presented in Fig.~\ref{Fig4} for the 4-terminal $\Psi$-shape geometry (depicted in the inset of Fig.~\ref{Fig3}a). Note that in order to obtain a significant effect of disorder, we used extremely (unrealistic) high values of disorder, much higher than the SO gap itself. A first general qualitative conclusion is that the ballistic results presented above are extremely robust with respect to all kinds of disorder. More precisely, we find, as expected, that extremely large values of $V_s$ or $V_z$ are needed to significantly affect the transport coefficients. ``Magnetic" impurities ($V_m$), which can induce intra-edge backscattering, do affect the QSH phase at weaker values. We note, however, that the system remains topological (i.e. does not become a simple ordinary insulator) as transport is still dominated by edge contributions: for instance, at $V_m=0.5$, the direct transmission probability $T_{20}$ (from lead 0 to lead 2) remains many orders of magnitude smaller than $T_{10}$ (from lead 0 to 1). The topological nature of the phase is encoded in the $\mathbb{Z}_2$ topological invariant, which cannot change unless the bulk gap is closed by the perturbations we consider. We checked (not shown) that the above results are essentially unaffected upon changing the SO gap by a factor of $2$, putting all three sorts of disorder simultaneously or changing the Fermi energy (inside the SO gap).

To rule out any concerns raised by the intra-edge scattering due to $V_m$, the corresponding intra-edge mean free path $l_e$ (or equivalently the localization length, as both are roughly equal for one-dimensional states) can be estimated using Fermi's golden rule: $l_e = v_d \tau_e$ where $v_d=\hbar^{-1}(dE/dk)$ is the drift velocity of the edge states (extracted for instance from Fig.~\ref{Fig2}), and $\hbar/\tau_e \propto (V_m)^2 |dk/dE|$. Alternatively, one may extract it directly from the numerical calculations of Fig.~\ref{Fig4}, since $T_{10}\approx 1- (L/l_e)$ (with $L$ the length along the edge, roughly $5$ nm in this instance). Fitting the small $V_m$ regime with $T_{10}\approx 1-3.0 V_m^2$, we obtain (in nm)
\begin{equation}
l_e \approx (V_m/t)^{-2} \; ,
\end{equation}
where we have explicitly restored the hopping amplitude $t$ (our energy unit) in order to get numbers. The ballistic results are essentially unaffected for systems smaller than $l_e$, while the edge states become localized for larger systems. In the context of a QSH phase induced by adatoms (using for instance the indium atoms proposed by Weeks et al. \cite{Weeks11}), a possible source of $V_m$-like disorder comes from the SO coupling induced by the adatoms themselves. Typical values for $V_m$ are smaller than 1 meV, which translates into extremely large intra-edge mean free paths $l_e > 1$ mm. Hence we estimate that this perturbation should be largely irrelevant in real size samples.

To summarize, all of our results are essentially unaffected by the presence of disorder, except when the Fermi energy lies in the vicinity of the QSH-QH transition, in which case strong disorder can give rise to a random network of QH and QSH regions through which a percolating cluster connecting opposite edges can therefore lead to backscattering\,\footnote{This effect will be all the more potent for zigzag-terminated ribbons, as they feature energy-dependent QSH edge states, whose width is expected to strongly increase when the energy approaches the SO gap \cite{Metalidis11}.}.

\section{Topological heterojunction}
\label{sec:Het}

We take advantage of the above described topological quantum phase transition as a function of the Fermi energy to propose a setup which allows for a direct junction between two different topological phases in the same sample.
Let us consider the case in which an additional electrostatic gate enables to split the system in two parts, one in which the Fermi level is in the QSH phase, and the other in which the Fermi level is in the QH phase (Fig.~\ref{Fig1}c). This constitutes a QSH-QH junction which shares some similarities with quantum Hall $n$-$p$ junctions previously fabricated in graphene \cite{Williams07,Ozyilmaz07}. Indeed, the incoming ``unhappy" spin at the junction has no other choice but to propagate along the interface in order to reach the only other available channels which lie on the opposite edge. This is reminiscent of the situation both spin channels have to face in the QH regime when they must cross a $n$-$p$ junction, since their direction of propagation on a given edge is reversed for negative energies. Various theoretical models have been proposed in the latter setup \cite{Abanin07,Tworzydlo07,Akhmerov08vv,Carmier11} but they all fail to explain the experimental observations \cite{Williams07,Ozyilmaz07}, probably due to some dephasing mechanism taking place in the vicinity of the Dirac point which is obscured by charge-density fluctuations (so-called electron-hole puddles). The system we consider could therefore provide a new perspective to solve this puzzle, as the QSH-QH transition takes place at a value of energy which can be far away from the Dirac point\,\footnote{Far from the Dirac point, however, the adatoms which induce the QSH phase in graphene \cite{Weeks11} may give rise to parasitic \cite{Kane05bis} Rashba-type SO couplings.} for realistic values of SO-induced QSH gap.

\begin{figure}[]
\begin{center}
\includegraphics[angle=0,width=1.0\linewidth]{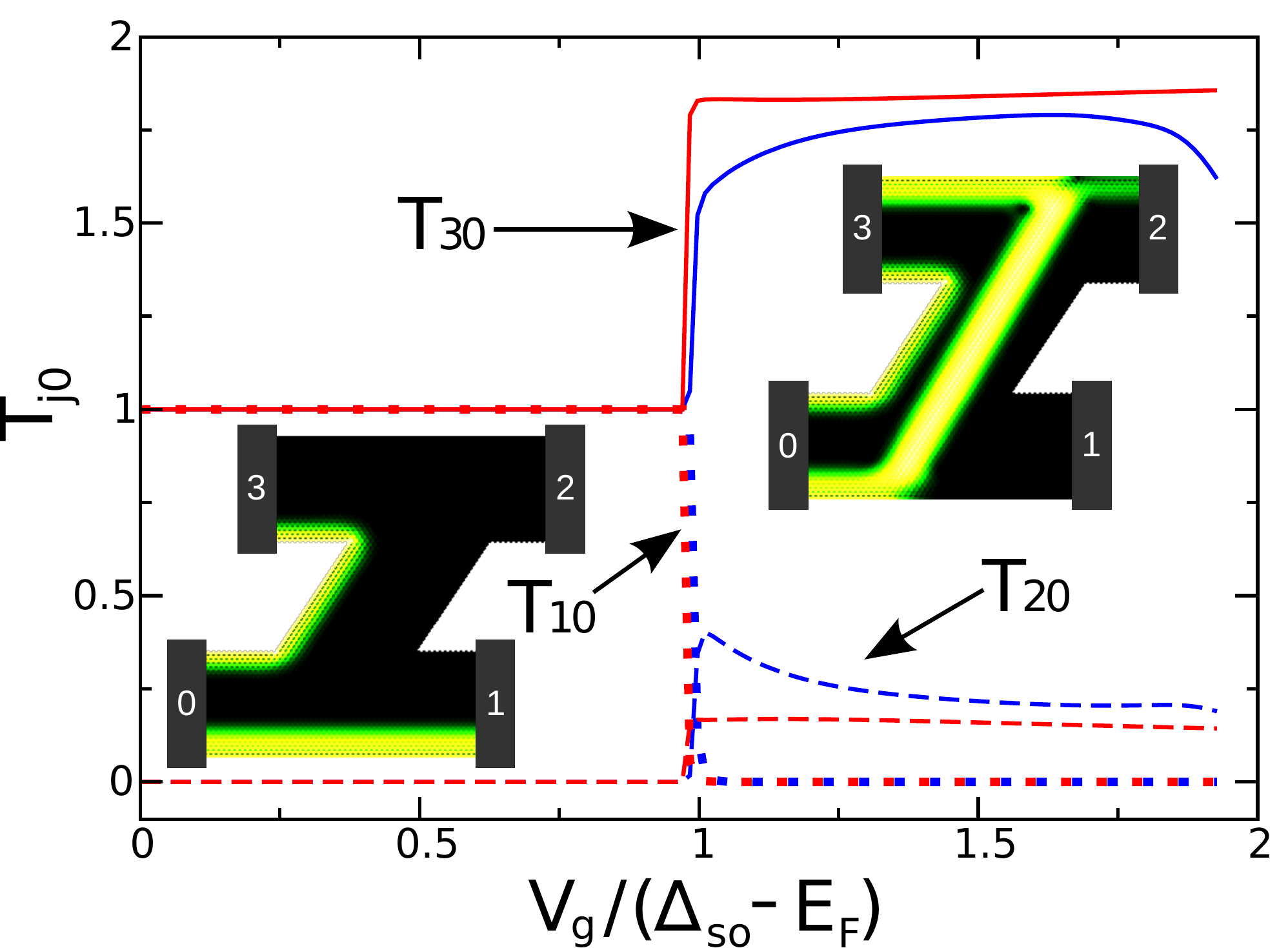}
\caption{Heterostructure in a 4 terminal Z-shape sample as depicted in Fig.~\ref{Fig1}c. Transmission probabilities from lead 0, where the current is injected, to outgoing leads 1 (dotted lines), 2 (dashed lines) and 3 (full lines) as a function of the top gate voltage $V_g$. When $V_g<(\Delta_{\text{so}}-E_F)$, such that left and right regions are in the QSH phase, current is perfectly transmitted by the QSH edge states, as shown in the current-density plot in the left inset. When $V_g$ is high enough ($V_g>(\Delta_{\text{so}}-E_F)$) that the right part of the sample enters the QH phase, a QSH-QH junction is created, which is characterized by a chiral state propagating along the interface. This is illustrated in the current-density plot shown in the right inset. Once it has reached the opposite edge, this chiral state is partially transmitted in lead 2, partially transmitted in lead 3, with proportions shown in the main plot. The light red curves correspond to an abrupt voltage change across the junction region while the dark blue curves correspond to a smooth transition. Parameters  are
$E_F/\Delta_{\text{so}}=0.58$ (in the left half of the sample), $\lambda_{\text{so}}=0.02$, $l_B\simeq8$, and widths $W_0=W_{\text{c}}=40\sqrt{3}$ for the leads and central region.}
\label{Fig5}
\end{center}
\end{figure}

More generally, our proposal offers the possibility of studying the nature of the state that propagates at the interface between QSH and QH phases, which are characterized by different topological invariants \cite{Thouless82,Kane05}. What we usually refer to as QSH (or QH) edge states are states propagating between QSH (or QH) insulators and a trivial insulator (the vacuum, typically). As the QSH insulator is characterized by a $\mathbb{Z}_2$ number, there is only one QSH topological phase: junctions between QSH phases with different Fermi energies (including $n$-$p$ junctions) have no effect on transport, as the spin-polarized states can propagate through these junctions. On the other hand, the QH topological invariant is a $\mathbb{Z}$ number, which counts the number of edge channels, and the notion of QH junctions therefore makes sense. In this case, one expects the existence of chiral propagating states, localized at the interface corresponding to the Landau level crossing. For QH $n$-$n'$ junctions, these states are ``bubbling" states \cite{Carmier11}, which simply follow the drifting Hall motion of charge carriers subjected to crossed electric and magnetic fields. For QH $n$-$p$ junctions, these states are ambipolar ``snake" states \cite{Williams11}, which can be seen as classical skipping orbits of mixed electron-``hole" character. The characteristics of the state propagating at the interface between QSH and QH insulators, on the other hand, are still unclear as far as we know.

Our proposal is the following. Consider the Z-shape geometry shown in Fig.~\ref{Fig1}c. It is split into two regions: the left half and the right half. A top gate is applied on the latter in order to tune the Fermi level in both regions independently. In Fig.~\ref{Fig5}, we plot, for a given value of the Fermi energy in the left half inside the QSH gap, the differential conductance $T_{i0}$ from lead 0 to leads 1, 2 and 3 as a function of the gate voltage (which determines the value of the Fermi energy in the right half). While the Fermi energy in the right half remains below the value of $\Delta_{\text{so}}$, transport characteristics remain unaffected by the gate (left inset of Fig.~\ref{Fig5}). However, as soon as the Fermi energy in the right half crosses the QSH gap, it gives rise to a QSH-QH junction characterized by a topological state at its interface (see right inset of Fig.~\ref{Fig5}). This chiral state propagates along the interface until it reaches the opposite edge, and then gets partially transmitted in lead 2, partially transmitted in lead 3, with proportions which depend on the microscopic details of the model (Fermi energies, the length of the interface, the smoothness of the potential step, the amount of disorder, the possible valley-space polarization of the edge states, etc), the study of which is left to subsequent work. This system constitutes a very efficient spin-polarized charge-current switching mechanism, as the current in lead 1 (respectively 2) can be reversibly switched from one (respectively zero) to zero (respectively non-zero) while simultaneously being spin-polarized (see Fig.~\ref{Fig5}). Additionally, this switching can be activated by simply tuning the voltage in the top gate over a very small energy range (whose lower bound will depend on the magnitude of the disorder in the vicinity of the transition), and should yield extremely sharp transitions with values of on/off current topologically protected from the presence of disorder\,\footnote{Note that a similar switching mechanism (without the spin polarization) could also be realized using a graphene quantum Hall $n$-$p$ junction, provided the charge-density fluctuations close to the Dirac point can be reduced.}.

\section{Conclusion}
\label{sec:Conc}

In summary, we showed that the transition between QSH and QH phases in graphene is independent of the value of the magnetic field (neglecting the weak effect of Zeeman splitting) and can be crossed simply by tuning the value of the Fermi energy across the SO-induced QSH gap. In a heterojunction, one of the spin species gives rise to a chiral state propagating along the interface between QSH and QH phases. The nature of this special state is a fascinating issue which could bring new light concerning the unresolved mystery of conductance plateaus observed in quantum Hall $n$-$p$ junctions \cite{Williams07,Ozyilmaz07}. We also showed that the tunable transition between the QSH and QH topological phases could serve as a spin-polarized charge-current switch with potentially extremely high, topologically protected, on/off ratios.
An interesting future direction of research could be to investigate whether this tunable topological phase transition can arise in bilayer graphene \cite{Novoselov06}, which also possesses zero-energy Landau levels \cite{McCann06} and has very recently been shown to host a $\mathbb{Z}_2$ topologically insulating phase \cite{Qiao12} for strong enough Rashba SO coupling.

\section*{Appendix: magnetic field in multi-terminal calculations}

In this appendix, we explain how to compute the Peierls phase $\phi_{ij}$ between two atoms $i$ and $j$ in multi-terminal systems \cite{Baranger89}. It is given by
\be
\phi_{ij}=\int_{{\bf r}_j}^{{\bf r}_i} {\bf A} \cdot d{\bf r}
\ee
where ${\bf r}_i=(x_i,y_i)$ is the spatial position of site $i$. A common choice for ${\bf A}$ is the Landau gauge (Lg)
\be
{\bf A}_{Lg}(x,y)= - By
\left(\begin{array}{c} 1 \\ 0 \end{array}\right)
\ee
which leads to
\be
\label{Lg}
\phi_{ij}^{Lg}=-B (x_i-x_j)\frac{y_i+y_j}{2}
\ee
(using linear paths between atoms). The numerical prescription follows simply: one calculates the real coordinates of the two atoms and uses the above equation to get $\phi_{ij}$.
An important aspect of the Landau gauge is that the phase depends on the $x$-coordinate only through the {\it difference} of $x$ between the two atoms $i$ and $j$. This is crucial for taking magnetic field into account in the leads: the leads are semi-infinite periodic systems made of layers. They are described by an intra-layer Hamiltonian $H_0$ and  an inter-layer Hamiltonian $V$. Within the Landau gauge, we find that the matrices $H_0$ and $V$ of {\it horizontal} leads are layer-independent, which allows the use of standard schemes to calculate their self-energies. However, general samples (such as the $\Psi$-shaped sample studied in this article) can have leads with an arbitrary angle $\theta$ with respect to the $y$-axis. For those leads, the corresponding gauge choice is
\be
{\bf A}({\bf r})= - B ({\bf r} \cdot {\bf e}_2) {\bf e}_1
\ee
with
\be
{\bf e}_1=\left(\begin{array}{c} \cos\theta \\ \sin\theta \end{array}\right) \; , \;
{\bf e}_2=\left(\begin{array}{c} -\sin\theta \\ \cos\theta \end{array}\right) \; , \;
\ee
which leads to
\be
\label{phi}
\phi_{ij}=\phi_{ij}^{Lg} + \Phi_i -\Phi_j \; ,
\ee
where the potential
\be
\label{toto}
\Phi_i= B (1-\cos 2\theta )\frac{x_i y_i}{2}  +B \sin 2\theta \frac{x_i^2 -y_i^2}{4}
\ee
is a pure gauge potential allowing to go from one choice of gauge to the other. A (possible) general prescription for an arbitrary system is now the following: one assigns a potential $\Phi_i=0$ to all sites except those belonging to a lead. For the latter, one uses Eq.~(\ref{toto}) with the appropriate angle $\theta$. Then one calculates the phases $\phi_{ij}$ according to Eq.~(\ref{phi}).

All the prescriptions above are given in real space. It is of course possible to calculate analytically the equivalent prescriptions in terms of the integer coordinates on the Bravais sublattices, as both are in one to one correspondance. However, for numerical purposes, it is more convenient to calculate the real space positions of the atoms
numerically and then use Eqs.~(\ref{Lg},\ref{toto},\ref{phi}) in order to use a lattice-independent prescription.

\begin{acknowledgments}
This work was supported by STREP ConceptGraphene, EC Contract ERC MesoQMC and ANR grant 2010-IsoTop.
\end{acknowledgments}


\end{document}